\documentclass[final,5p,times,twocolumn,authoryear]{elsarticle}

\usepackage[authoryear]{natbib}
\usepackage{graphicx}
\usepackage{subcaption}
\usepackage{amssymb,amsmath}
\usepackage{pdflscape}
\usepackage{adjustbox}
\usepackage{hyperref}
\usepackage{booktabs}
\usepackage{lipsum}
\usepackage{epsfig}
\usepackage{xcolor}
\usepackage{tabularx}
\usepackage{footnote}
\usepackage{soul}
\usepackage{multirow}
\usepackage{multicol}
\usepackage[T1]{fontenc}
\usepackage{mathpazo}
\usepackage{lineno,hyperref}
\usepackage{textcomp}
\usepackage{url}

\newcommand{\revone}[1]{#1}
\newcommand{\revtwo}[1]{#1}

\usepackage[utf8]{inputenc}
\usepackage{booktabs}

\usepackage{amsthm}

\journal{High Energy Astrophysics}

\begin{document}

\begin{frontmatter}



\title{Exploring the Transitional Parameter Space of Blazars using Gamma-ray and X-ray Population Diagnostics}


%

\author{Zahoor Malik\corref{cor1}\fnref{label1}}
\ead{malikzahoor313@gmail.com}
\affiliation[label1]{Department of Physics, University of the Free State, Bloemfontein, 9301, South Africa}
\cortext[cor1]{Corresponding author: Zahoor Malik}

\author{Sikandar Akbar\fnref{label2}}
\ead{darprince46@gmail.com}
\affiliation[label2]{Department of Physics, University of Kashmir, Srinagar 190006, India}

\author{Zahir Shah\fnref{label3}}
\ead{shahzahir4@gmail.com}
\affiliation[label3]{Manipal Centre for Natural Sciences, Centre of Excellence, Manipal Academy of Higher Education, Manipal - 576104, India}

\author{B. van Soelen\fnref{label1}}
\ead{vansoelenb@ufs.ac.za}

\begin{abstract}
We investigate the $\gamma$-ray and X-ray population properties of changing-look blazars (CLBs) using sources from the Fourth \textit{Fermi} LAT Source Catalog Data Release 4 (4FGL-DR4) together with X-ray information from the Living \textit{Swift} XRT Point Source (LSXPS) catalog. The CLB sample is compared with large populations of confirmed BL Lac objects (BLLs) and flat-spectrum radio quasars (FSRQs) using spectral, variability, and broadband properties. In the $\gamma$-ray parameter space, CLBs mainly occupy intermediate and overlap regions between the BLL and FSRQ populations. However, the centroid locations in different parameter planes, along with the PCA and UMAP projections, show that the CLB population lies closer to the FSRQ region. The X-ray analysis also shows a similar behavior, where the overall distribution of CLBs in the X-ray parameter space is found to be nearer to FSRQs than to BLLs. In addition, the X-ray/$\gamma$-ray coupling relations and random-forest classification probabilities are consistent with this trend. Overall, the results suggest that CLBs form a transitional population between the two main blazar subclasses while retaining characteristics closer to the FSRQ population.
\end{abstract}



\end{frontmatter}


\section{Introduction}
\label{sec:intro}

Blazars are a subclass of radio-loud active galactic nuclei (AGNs) whose relativistic jets are oriented close to the observer's line of sight \citep{blandford1978pittsburgh,1995PASP..107..803U}. Their observed emission is strongly Doppler boosted and is dominated by non-thermal radiation extending from radio wavelengths to $\gamma$ rays. Blazars are characterized by rapid flux variability, high optical polarization, compact radio structure, and apparent superluminal motion \citep{2016ARA&A..54..725M,2019ARA&A..57..467B}. Based on the strength of optical emission lines, blazars are traditionally classified into two main subclasses: flat-spectrum radio quasars (FSRQs), which show prominent broad emission lines, and BL Lacertae objects (BLLs), which exhibit weak or nearly absent emission lines with equivalent widths smaller than 5~\r{A} \citep{1991ApJ...374..431S,1991ApJS...76..813S}. This observational distinction is generally associated with differences in accretion efficiency, jet power, and the surrounding external radiation environment \citep{1998MNRAS.301..451G,2016Galax...4...36G}. FSRQs typically possess softer $\gamma$-ray spectra and stronger long-term variability, whereas BLLs generally show harder spectra and weaker external radiation fields \citep{1998MNRAS.299..433F,2011ApJ...740...98M,2015ApJ...810...14A,2020ApJ...892..105A}.

Although the BLL--FSRQ classification scheme describes most blazars, an increasing number of sources have been observed to transition between the two subclasses. These systems, commonly known as changing-look or transitional blazars (CLBs), exhibit significant changes in their observed properties, leading to transitions between FSRQ-like and BLL-like states (\citealt{2015ApJ...800..144L,2016ApJ...826..188R,2016AJ....151...32A,Mishra_2021,2021AJ....161..196P} and references therein). Similar to changing-look AGNs, CLBs challenge the idea of rigid subclass boundaries and may provide important insight into the physical connection between different blazar populations \citep{2023NatAs...7.1282R}. Recent multiwavelength studies and large survey observations have substantially expanded the known sample of candidate CLBs (\citealt{Mishra_2021,2021AJ....161..196P,2021Univ....7..372F,2022Univ....8..587F,Xiao_2022,2024ApJ...962..122K} and references therein).

The physical mechanism responsible for the changing-look behavior in blazars is still not fully understood. Proposed explanations include strong variability of the relativistically beamed jet continuum, changes in the accretion state, variations in the jet bulk Lorentz factor, or transitions in the disk--jet coupling process \citep{1995ApJ...452L...5V,2009A&A...496..423B,2012MNRAS.420.2899G,2014ApJ...797...19R,2019RNAAS...3...92P}. In some cases, strong non-thermal jet emission may temporarily overwhelm the broad-line region, causing an FSRQ to appear observationally similar to a BL Lac object \citep{2012MNRAS.425.1371G}. Observational effects such as redshift, spectral coverage, signal-to-noise ratio, and spectral resolution may also influence the optical classification of some sources \citep{2015MNRAS.449.3517D,2021AJ....161..196P}. These systems therefore provide a useful opportunity to investigate the relation between jet emission, accretion processes, and optical spectral classification in blazars.

Large $\gamma$-ray surveys have enabled detailed population studies of blazars and their high-energy properties \citep{2020ApJS..247...33A,2020ApJ...892..105A}. Previous works have shown that parameters such as $\gamma$-ray photon index, spectral curvature, variability, and broadband properties provide important distinctions between BLLs and FSRQs \citep{2016MNRAS.462.3180C,2017A&A...602A..86L,2020MNRAS.493.1926K,2022JCAP...04..023B,2023ApJ...946..109A,61tz-jk8c}. Recent studies based on $\gamma$-ray population analyses further suggest that CLBs occupy intermediate regions between the canonical blazar subclasses \citep{2024ApJ...962..122K,10.1093/mnras/stag542}. However, most previous investigations have mainly focused on optical changing-look behavior, source identification, or subclass classification, while the multidimensional population properties of CLBs in combined $\gamma$-ray and X-ray parameter spaces remain relatively unexplored.

In this work, we study the population properties of CLBs using $\gamma$-ray information from the 4FGL-DR4 catalog \citep{2020ApJS..247...33A, 2023arXiv230712546B} together with X-ray data from the LSXPS catalog \citep{2023MNRAS.518..174E}. We compare the CLB population with confirmed BLLs and FSRQs using their spectral, \revone{variability (quantified by the \textit{Fermi}-LAT Variability Index)}, and broadband properties. In addition to conventional statistical comparisons, we employ principal component analysis (PCA; \citealt{Jolliffe2002}), Uniform Manifold Approximation and Projection (UMAP; \citealt{2018arXiv180203426M}), and random-forest (RF) classification techniques \citep{2001MachL..45....5B} to investigate the multidimensional topology of the blazar population. The primary aim of this work is to examine whether CLBs occupy statistically intermediate regions between the two established blazar subclasses and to determine the extent to which their overall population properties are associated with either BLLs or FSRQs.

\revone{This paper is organized as follows. In Section~\ref{sec:data}, we describe the sample selection procedure and the construction of the $\gamma$-ray and X-ray datasets. Section~\ref{sec:stats} presents the $\gamma$-ray population analysis, including the univariate and bivariate statistical distributions, the multidimensional population structure based on PCA and UMAP, and the random-forest classification. The X-ray population analysis and X-ray/$\gamma$-ray coupling results are presented in Section~\ref{sec:xray}. Finally, the results are summarized and discussed in Section~\ref{sec:discussion}.}

\section{Sample Selection and Data Preparation}
\label{sec:data}

For the present analysis, we used the reported/confirmed changing-look blazar sample compiled in the online changing-look (transition) blazar catalog of \citet{shi_ju_kang_2023_8239456}. The catalog includes CLBs or transitional sources reported in previous studies, including \cite{1995ApJ...452L...5V};
\cite{2009A&A...496..423B};
\cite{2011MNRAS.414.2674G};
\cite{2012ApJ...748...49S};
\cite{2014ApJ...797...19R};
\cite{2014MNRAS.445.4316C};
\cite{2016AJ....151...32A};
\cite{2019RNAAS...3...92P};
\cite{2019MNRAS.484L.104P};
\cite{2021AJ....161..196P};
\cite{2021Univ....7..372F};
\cite{Mishra_2021};
\cite{2022ApJ...935....4Z};
\cite{2022Univ....8..587F};
\cite{Xiao_2022}. The $\gamma$-ray properties of the sources were obtained from the 4FGL-DR4 catalog \citep{2020ApJS..247...33A, 2023arXiv230712546B}. The CLB sample was crossmatched with the 4FGL-DR4 catalog using the associated source names, resulting in a sample of 109 confirmed CLBs with identified 4FGL counterparts. For comparison, samples of confirmed BLLs and FSRQs were selected from the 4FGL-DR4 catalog using the \texttt{CLASS1} classification field, where sources classified as \texttt{bll} and \texttt{fsrq} were assigned to the corresponding subclasses. Sources belonging to the confirmed CLB sample were excluded from the normal BLL and FSRQ populations in order to avoid duplication between subclasses.

\revone{The parent $\gamma$-ray sample consists of 2314 sources, including 109 CLBs, 1429 BLLs, and 776 FSRQs}. For each source, we considered the $\gamma$-ray photon index ($\Gamma$), pivot energy, variability index, and log-parabola curvature parameter ($\beta$). In addition, redshift information available from the associated 4LAC-DR3 catalog \citep{2022ApJS..263...24A} was incorporated for sources with measured redshifts, and the integrated $\gamma$-ray energy flux ($100~{\rm MeV}$--$100~{\rm GeV}$) was used to estimate the $\gamma$-ray luminosities in order to examine luminosity-related trends within the population distributions.

\begin{figure}
\centering
\includegraphics[width=0.95\linewidth]{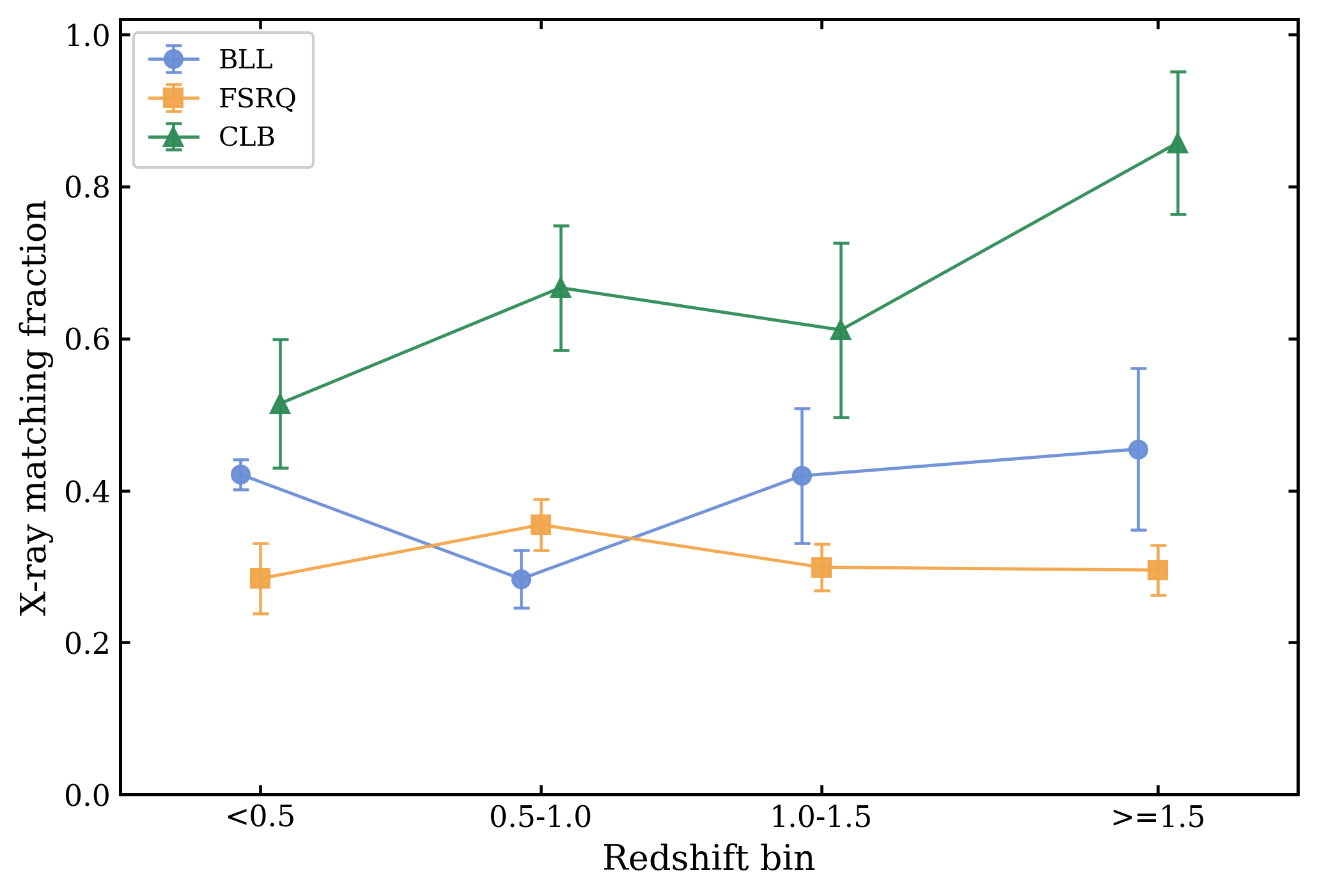}
\caption{\revone{X-ray matching fraction as a function of redshift for the CLB, BLL, and FSRQ populations. The matching fraction is defined as the ratio of LSXPS-associated sources to the parent 4FGL-DR4 sources with measured redshift in each common redshift bin. Error bars represent the corresponding binomial uncertainties.}}
\label{fig:xray_matching_redshift}
\end{figure}

To include the X-ray information, the $\gamma$-ray sample was crossmatched with the LSXPS catalog \citep{2023MNRAS.518..174E}. The positional matching between the 4FGL-DR4 and LSXPS catalogs was performed using the angular separation between source coordinates,

\begin{equation}
\cos\theta = \sin\delta_1 \sin\delta_2 + \cos\delta_1 \cos\delta_2 \cos(\alpha_1-\alpha_2),
\end{equation}

where $\alpha$ and $\delta$ represent the right ascension and declination of the two sources, respectively, and $\theta$ is the angular separation between them. Several matching radii between 10 and 180 arcsec were examined in order to balance sample completeness and contamination. A matching radius of 60 arcsec was therefore adopted for the final X-ray analysis, resulting in 887 associated sources consisting of 66 CLBs, 586 BLLs, and 235 FSRQs. \revone{To examine whether the LSXPS-associated X-ray sample is representative of the parent $\gamma$-ray population, we investigated the X-ray matching fraction as a function of redshift for each blazar subclass. The matching fraction was computed in common redshift bins as the ratio of the number of LSXPS-associated sources to the number of parent 4FGL-DR4 sources with measured redshift. The resulting diagnostic is presented in Figure~\ref{fig:xray_matching_redshift}. In addition, we quantified the dependence of the matching probability on redshift using the Spearman rank correlation between the individual-source redshift and a binary X-ray matched flag. The CLB ($\rho=0.149$, $p=0.138$) and FSRQ ($\rho=-0.001$, $p=0.969$) populations show no statistically significant redshift dependence, whereas the BLL population exhibits only a weak negative correlation ($\rho=-0.127$, $p=3.2\times10^{-4}$). These results indicate that the X-ray matching probability remains broadly stable with redshift for the different populations, supporting the use of the LSXPS-associated sample for the subsequent X-ray population analysis.}

The X-ray quantities used in this work include the X-ray photon index, HR2 hardness ratio, and X-ray flux ($F_{\rm X}$). The X-ray flux was used together with the available redshift information to estimate the X-ray luminosities for the associated sources. \revone{Among the available hardness-ratio quantities in the LSXPS catalog, both HR1 and HR2 were examined during the exploratory analysis. The corresponding HR1 analysis is presented in~\ref{appendix:hr1}. Since HR2 provided comparatively clearer subclass separation and more stable statistical behavior, it was adopted as the primary hardness-ratio diagnostic throughout this work.} These parameters describe the X-ray spectral shape, spectral hardness, and broadband emission properties of the sources, providing a consistent basis for comparing the CLB, BLL, and FSRQ populations.

\section{Gamma-ray Properties}
\label{sec:stats}
\subsection{Univariate and Bivariate Statistical Analysis}
\label{sec:stats1}

\begin{figure*}
\centering

\includegraphics[width=0.85\textwidth]{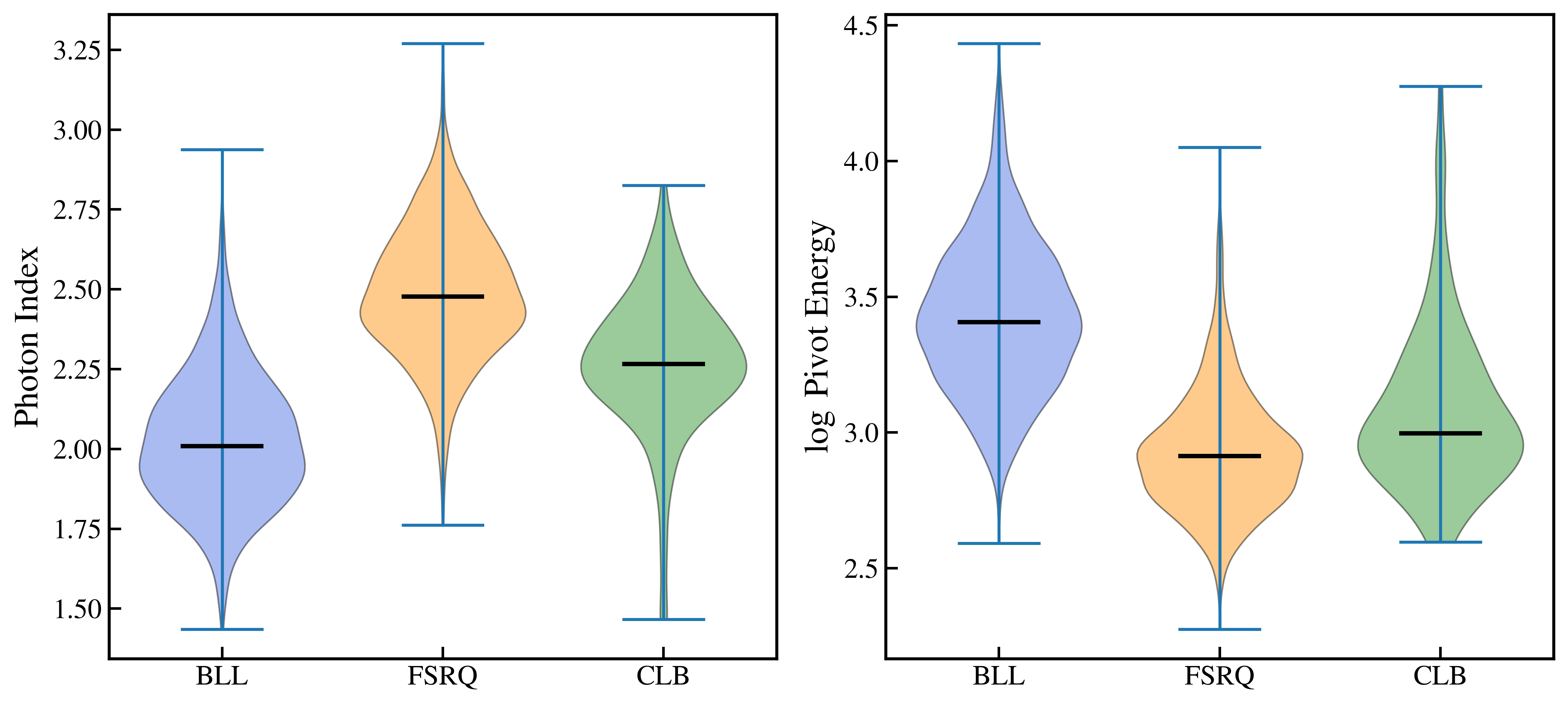}

\vspace{0.3cm}

\includegraphics[width=0.85\textwidth]{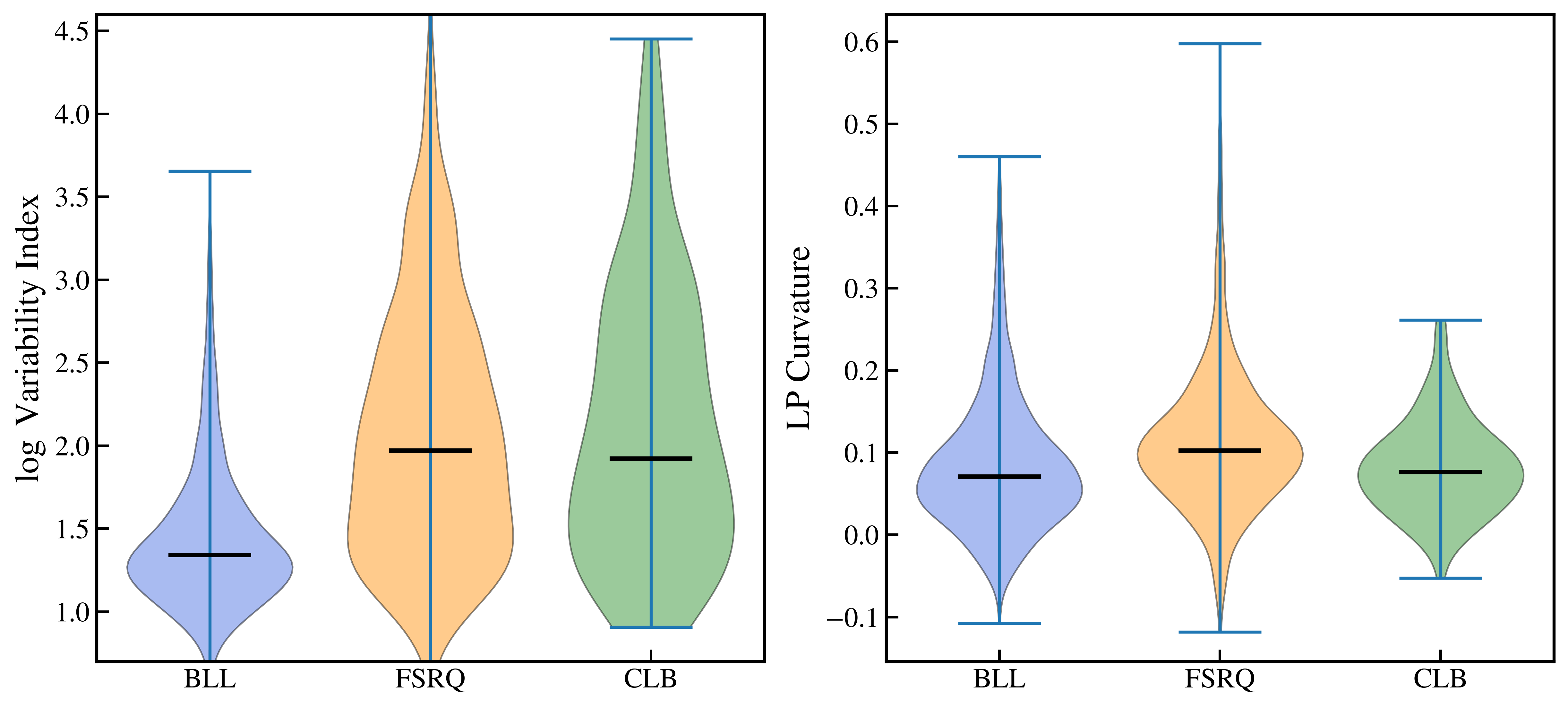}

\caption{
Violin distributions of the principal $\gamma$-ray parameters for the CLB, BLL, and FSRQ populations. 
The upper panels show the distributions of the photon index and logarithmic pivot energy, while the lower panels present the distributions of the logarithmic variability index and log-parabola (LP) curvature parameter ($\beta$). 
The violin distributions illustrate the median behavior, interquartile spread, and overall population morphology of the three subclasses. 
}
\label{fig:violin}
\end{figure*}

In order to investigate the $\gamma$-ray population properties of changing-look blazars, we performed a comparative statistical analysis between the CLB, BLL, and FSRQ samples using the principal $\gamma$-ray parameters extracted from the 4FGL-DR4 catalog. The analysis primarily focuses on the photon index ($\Gamma$), pivot energy, variability index, and log-parabola curvature parameter ($\beta$). The one-dimensional distributions of these parameters are presented in Figure~\ref{fig:violin}. Logarithmic transformations were applied to the pivot energy and variability index owing to their broad dynamical ranges.

The photon-index distributions reveal a clear separation between BLL and FSRQ populations. BLLs predominantly occupy harder spectral regimes characterized by lower photon-index values, whereas FSRQs are systematically shifted toward softer spectra. The CLB population mainly occupies intermediate and overlapping regions between the two subclasses. A similar trend is observed in the pivot-energy distributions, where BLLs occupy higher pivot-energy regimes while FSRQs are concentrated toward lower values. The CLBs again populate the overlap region between the two populations with a broader overall spread. The lower panels of Figure~\ref{fig:violin} present the distributions of the logarithmic variability index and the log-parabola curvature parameter. FSRQs generally exhibit stronger long-term variability compared to BLLs, while the CLB population shows intermediate variability behavior with substantial overlap toward the FSRQ regime. In contrast, the curvature parameter exhibits comparatively weaker separation between the subclasses, with significant overlap among all three populations.

To quantify the statistical differences between the populations, we performed two-sample Kolmogorov--Smirnov (KS) tests between the CLB sample and the canonical BLL and FSRQ populations. The resulting KS statistics and corresponding $p$-values are summarized in Table~\ref{tab:ks_gamma}. The photon-index and pivot-energy distributions show significant statistical separation between CLBs and BLLs, while the corresponding CLB--FSRQ separations, although still statistically significant, are comparatively smaller. A similar behavior is observed for the variability-index distributions, where the CLB--FSRQ comparison yields a relatively small KS statistic and a comparatively large $p$-value. These trends suggest that the overall $\gamma$-ray properties of CLBs are statistically more closely associated with the FSRQ population than with the BLL population. In contrast, the curvature parameter shows weaker discriminatory behavior between the subclasses.

\begin{table*}
\centering
\caption{Kolmogorov--Smirnov (KS) test results for the principal $\gamma$-ray parameters comparing the CLB population with BLL and FSRQ subclasses.}
\label{tab:ks_gamma}
\begin{tabular}{lcccc}
\hline
Parameter & KS$_{\rm CLB-BLL}$ & $p$-value & KS$_{\rm CLB-FSRQ}$ & $p$-value \\
\hline
Photon Index ($\Gamma$) & 0.566 & $9.58\times10^{-31}$ & 0.449 & $7.16\times10^{-18}$ \\

log Pivot Energy & 0.527 & $1.95\times10^{-26}$ & 0.269 & $1.58\times10^{-6}$ \\

log Variability Index & 0.445 & $1.47\times10^{-18}$ & 0.074 & $6.48\times10^{-1}$ \\

LP Curvature ($\beta$) & 0.091 & $3.56\times10^{-1}$ & 0.216 & $2.38\times10^{-4}$ \\
\hline
\end{tabular}
\end{table*}

\begin{figure*}
\centering

\includegraphics[width=0.92\textwidth]{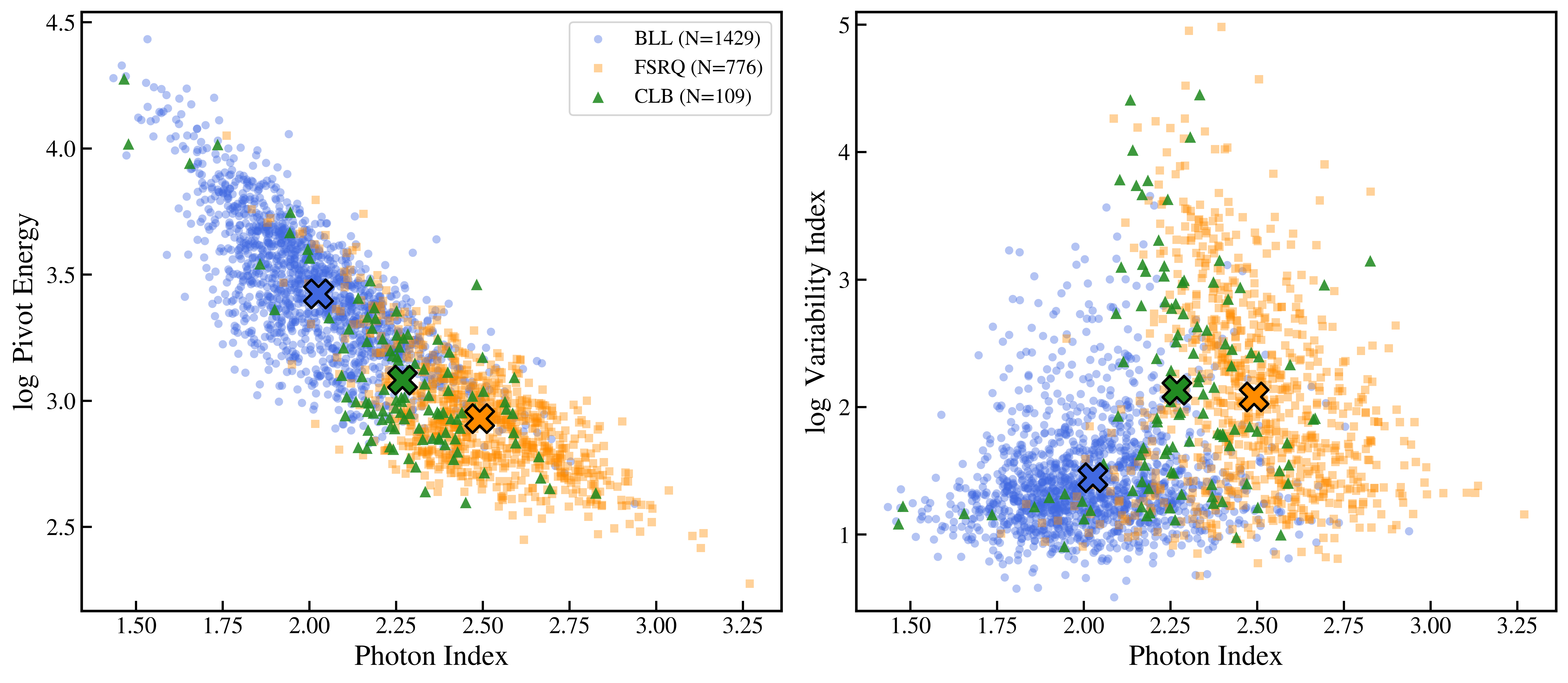}

\caption{
Representative two-parameter $\gamma$-ray distributions for the CLB, BLL, and FSRQ populations using combinations of photon index, pivot energy, and variability index. The corresponding centroid positions \revtwo{(arithmetic means)} are also shown in each panel.
}
\label{fig:scatter}
\end{figure*}

In addition to the one-dimensional distributions, we examined the location of CLBs in different two-parameter spaces. Representative scatter plots are shown in Figure~\ref{fig:scatter}. Among the explored parameter combinations, the $\Gamma$--pivot-energy plane shows the clearest separation between the two classical subclasses. BLLs mainly occupy the hard-spectrum and high-pivot-energy region, whereas FSRQs are concentrated toward steeper spectra and lower pivot energies. The CLB population is primarily distributed within the overlap region and forms a continuous connection between the BLL and FSRQ populations rather than appearing as a completely isolated group. A similar behavior is observed in the $\Gamma$--variability-index plane, where several CLBs extend into the FSRQ-dominated region while others remain comparatively closer to the BLL population, indicating that the CLB sample spans a broad range of spectral and variability properties. The distribution of CLBs in these parameter spaces is not symmetric between the two subclasses, with the centroid positions shifted comparatively closer to the FSRQ region. These trends suggest that although CLBs occupy intermediate and overlapping regions in the $\gamma$-ray parameter space, their overall population properties are comparatively closer to FSRQs. This motivates a more detailed multidimensional investigation of the population structure.

\subsection{Multidimensional Population Structure (PCA/UMAP)}
\label{sec:latent}

To further investigate the population structure of CLBs in the multidimensional $\gamma$-ray parameter space, we performed PCA and UMAP using the photon index, logarithmic pivot energy, logarithmic variability index, and log-parabola curvature parameter. \revone{Prior to the PCA and UMAP analyses, all parameters were standardized in order to avoid dominance from variables with larger numerical ranges (Note that the one-dimensional and two-dimensional analyses presented in Section~\ref{sec:stats1} did not require this standardization, as they treated each parameter independently).} \revone{Only sources with complete measurements for all selected parameters were retained for the multidimensional analysis. This resulted in an analysis sample of 108 CLBs, 1429 BLLs, and 775 FSRQs, with one CLB and one FSRQ excluded because of missing measurements for one or more of the selected $\gamma$-ray parameters.}

\revtwo{The multivariate analyses were carried out in \textsc{Python} using
\texttt{scikit-learn} 1.8.0 \citep{scikit-learn} and \texttt{umap-learn} 0.5.12.
The principal algorithm configurations are summarized in Table~\ref{tab:hyper}.
Two principal components were retained, accounting for 54.5 and 25.1 per cent of
the total variance respectively (79.7 per cent cumulatively). The UMAP embedding
used \texttt{n\_neighbors} = 25, \texttt{min\_dist} = 0.15, two output dimensions
and the Euclidean metric, and a fixed random seed of 42 was adopted throughout.
Because the geometry of a UMAP embedding can depend on the choice of
neighborhood hyperparameters, we repeated the projection over a grid of
\texttt{n\_neighbors} between 5 and 100 and \texttt{min\_dist} between 0.0 and
0.5. The qualitative structure is unchanged in all 30 configurations examined,
and the CLB centroid remains closer to the FSRQ centroid than to the BLL centroid
in every case, with the ratio of the two centroid distances ranging between 2.4
and 6.2. The random-forest classifier used 500 trees with a maximum depth of 5, a
minimum of 10 sources per leaf, and balanced class weights, and its performance
was evaluated using five-fold stratified cross-validation.}

\begin{table*}
\centering
\caption{\revtwo{Configuration of the multivariate algorithms used in this work.
A common random seed of 42 was used for all stochastic components.}}
\label{tab:hyper}
\begin{tabular}{llc}
\hline
Analysis & Setting & Value \\
\hline
Preprocessing & Standardization & zero mean, unit variance \\
\hline
PCA           & Components retained & 2 \\
              & Explained variance  & 54.5\%, 25.1\% \\
\hline
UMAP          & \texttt{n\_neighbors} & 25 \\
              & \texttt{min\_dist}    & 0.15 \\
              & Output dimensions     & 2 \\
              & Metric                & Euclidean \\
\hline
Random forest & Number of trees        & 500 \\
              & Maximum depth          & 5 \\
              & Minimum sources / leaf & 10 \\
              & Class weighting        & balanced \\
              & Cross-validation       & 5-fold stratified \\
\hline
\end{tabular}
\end{table*}

\begin{figure*}
\centering

\includegraphics[width=0.8\textwidth]{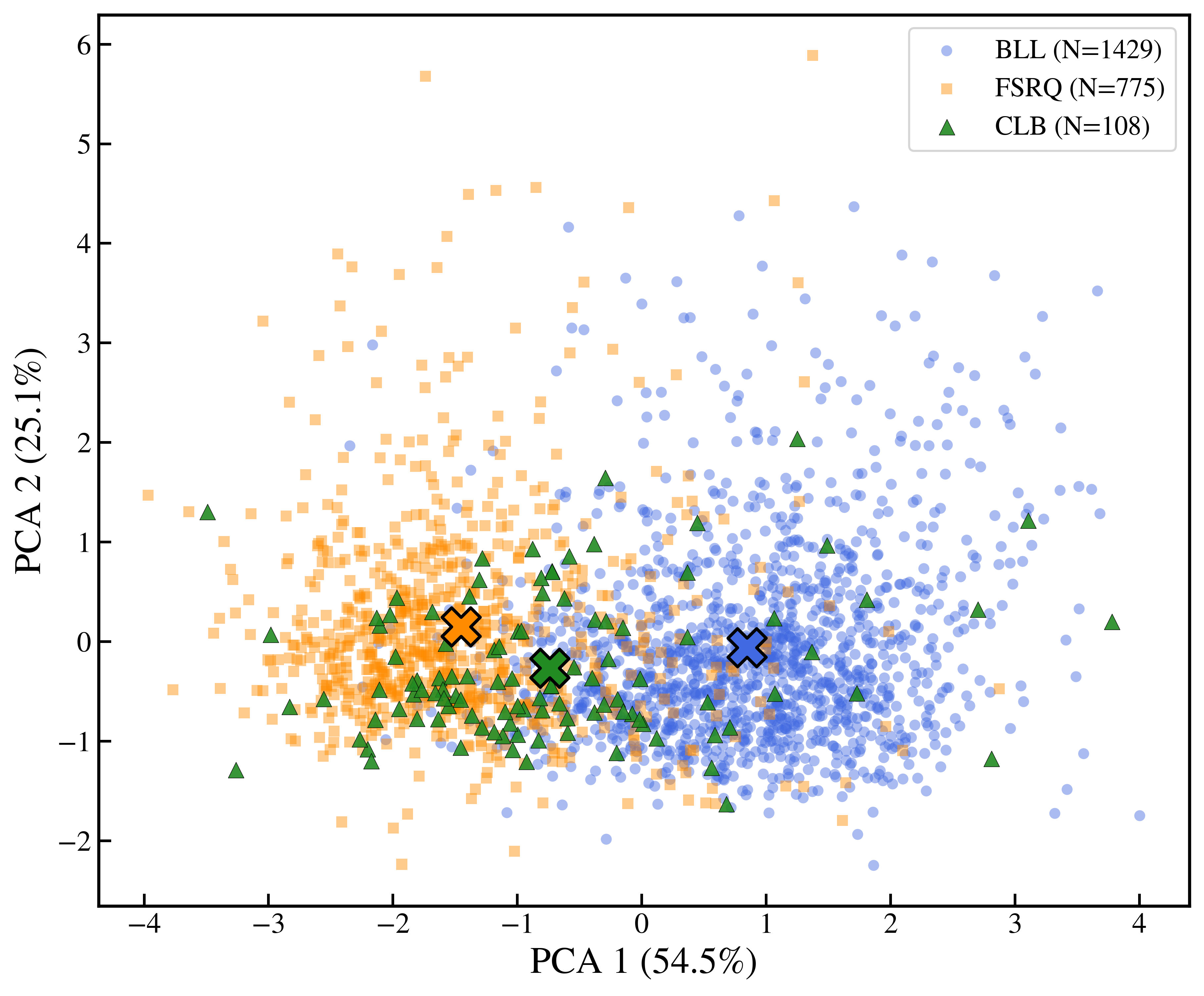}

\caption{
Two-dimensional principal component analysis (PCA) projection of the CLB, BLL, and FSRQ populations using the selected $\gamma$-ray parameters.
}
\label{fig:pca}
\end{figure*}

The two-dimensional PCA projection is shown in Figure~\ref{fig:pca}. The distribution reveals a noticeable separation between BLL and FSRQ populations, with the two subclasses predominantly occupying different regions of the PCA space. The CLB population mainly occupies the overlap region between the two subclasses rather than forming a completely separate cluster. Several CLBs extend into the BLL-dominated region, whereas a considerable fraction is located within or near the FSRQ region. The centroid position of the CLB population is also shifted comparatively closer to the FSRQ region. To quantify this trend, we calculated the Euclidean distances between the subclass centroids, each defined as the arithmetic mean position of the corresponding population, in the two-dimensional PCA projection. The CLB centroid lies closer to the FSRQ centroid (distance = 0.825) than to the BLL centroid (distance = 1.598), quantitatively confirming the preferential location of the CLB population toward the FSRQ region in the PCA space. The parameter contributions to the principal components are summarized in Table~\ref{tab:pca}. The first principal component is mainly influenced by the photon index and pivot energy, with an additional contribution from the variability index. In contrast, the second principal component is dominated primarily by the curvature parameter ($\beta$). \revone{Compared with the first principal component, the weaker subclass separation along the second principal component suggests that spectral curvature alone is less effective in distinguishing the different blazar populations.}

\begin{table}
\centering
\caption{Principal component loadings for the $\gamma$-ray parameters used in the PCA analysis.}
\label{tab:pca}
\begin{tabular}{lcc}
\hline
Parameter & PCA1 & PCA2 \\
\hline
PL Index & $-0.596$ & $0.070$ \\
log Pivot Energy & $0.654$ & $0.030$ \\
LP Curvature ($\beta$) & $-0.064$ & $0.980$ \\
log Variability Index & $-0.461$ & $-0.184$ \\
\hline
\end{tabular}
\end{table}

\begin{figure*}
\centering

\includegraphics[width=0.8\textwidth]{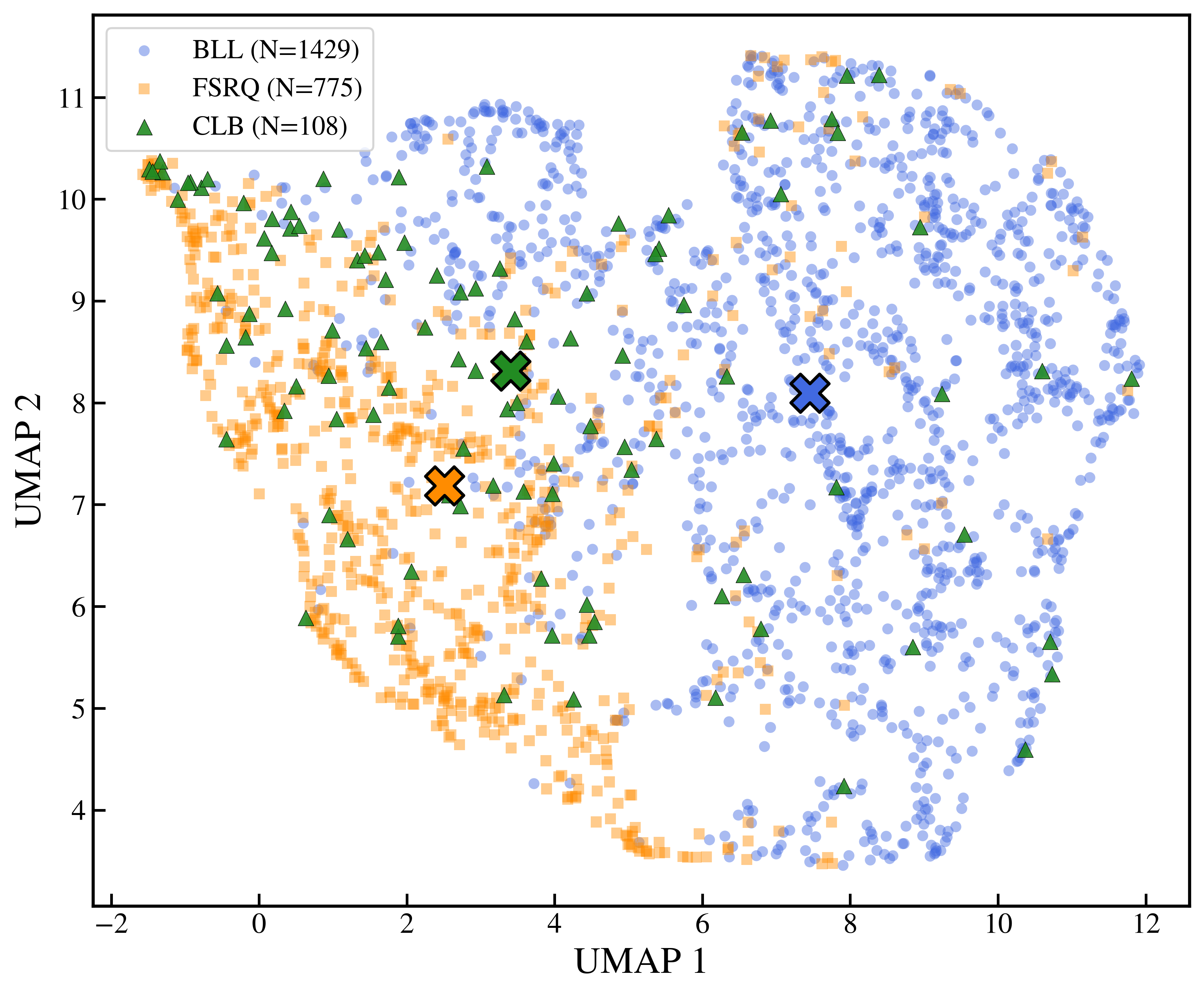}

\caption{
Two-dimensional UMAP projection of the CLB, BLL, and FSRQ populations derived from the selected $\gamma$-ray parameters. 
}
\label{fig:umap}
\end{figure*}

The corresponding UMAP projection is presented in Figure~\ref{fig:umap}. Similar to the PCA results, the BLL and FSRQ populations occupy comparatively distinct regions in the UMAP space, while the CLBs are mainly distributed across the intermediate region connecting the two subclasses rather than forming a completely isolated group. While a fraction of CLBs overlaps with the BLL-dominated region, a comparatively larger number of sources are distributed within or around the FSRQ region. The centroid position of the CLB population in the UMAP space is again shifted comparatively closer to the FSRQ population. Consistent with the PCA results, the Euclidean distance between the CLB and FSRQ centroids is 1.438, whereas the corresponding CLB--BLL centroid distance is 4.057, indicating that the CLB population remains closer to the FSRQ population in the nonlinear UMAP projection.

Overall, both the PCA and UMAP analyses show that CLBs occupy intermediate and overlapping regions between the BLL and FSRQ populations in the multidimensional $\gamma$-ray parameter space. However, the centroid behavior and overall distribution trends consistently indicate that the CLB population is comparatively closer to the FSRQ population.

\revtwo{As an independent check that does not use the source classifications, we additionally performed an unsupervised three-component Gaussian mixture analysis of the same parameter space. The CLB population is not recovered as a distinct component, and the sources are instead distributed among the components occupied by the canonical subclasses. The analysis and its relation to previous work are presented in~\ref{appendix:gmm}.}

\subsection{Machine-learning Classification}
\label{sec:ml}

\revone{As an additional test of the population behavior of CLBs, we performed an RF classification analysis using the same 4FGL-DR4 $\gamma$-ray analysis sample employed in the PCA and UMAP analyses (108 CLBs, 1429 BLLs, and 775 FSRQs).} The classifier was trained exclusively on the confirmed BLL and FSRQ populations using the photon index, logarithmic pivot energy, logarithmic variability index, and log-parabola curvature parameter as input features. To account for the class imbalance between the BLL and FSRQ populations, class weights were applied during the training process. The CLB sample was excluded from the training stage and used only for the final probability analysis.

The RF classifier shows good performance in separating the canonical subclasses, with a cross-validation accuracy of $0.898 \pm 0.011$ and a ROC--AUC score of $0.956 \pm 0.009$, as summarized in Table~\ref{tab:rf_summary}. To further understand which parameters drive the RF classification, we examined the feature importance of the input variables. The classification is primarily driven by the photon index (59.2\%) and logarithmic pivot energy (30.2\%), while the logarithmic variability index (8.4\%) and log-parabola curvature parameter (2.2\%) contribute comparatively less. This indicates that the classification is driven predominantly by the spectral properties of the sources, rather than by their long-term variability. \revone{Consequently, the predicted probabilities for CLBs should not be interpreted as calibrated class-membership probabilities. Rather, they quantify the relative similarity of each CLB to the regions of parameter space occupied by the two canonical subclasses.}

\begin{table*}
\centering
\caption{Summary of the random-forest (RF) classification analysis performed using the BLL and FSRQ populations. The classifier was trained using the photon index, pivot energy, variability index, and spectral-curvature parameters. The table additionally summarizes the resulting FSRQ probability distribution for the CLB population. \revone{The reported probabilities represent similarity to the canonical training populations rather than calibrated class-membership probabilities.}}
\label{tab:rf_summary}
\begin{tabular}{lc}
\hline
Quantity & Value \\
\hline

Training sample size & 2204 \\
BLL training sources & 1429 \\
FSRQ training sources & 775 \\

Cross-validation accuracy & $0.898 \pm 0.011$ \\
Cross-validation ROC-AUC & $0.956 \pm 0.009$ \\

Mean $P({\rm FSRQ})$ for CLBs & 0.58 \\
Median $P({\rm FSRQ})$ for CLBs & 0.66 \\

Number of CLBs with $P({\rm FSRQ}) > 0.5$ & 64 \\
Number of CLBs with $P({\rm FSRQ}) > 0.7$ & 50 \\
Number of CLBs with $P({\rm FSRQ}) > 0.9$ & 26 \\

\hline
\end{tabular}
\end{table*}

\begin{figure*}
\centering
\includegraphics[width=0.62\textwidth]{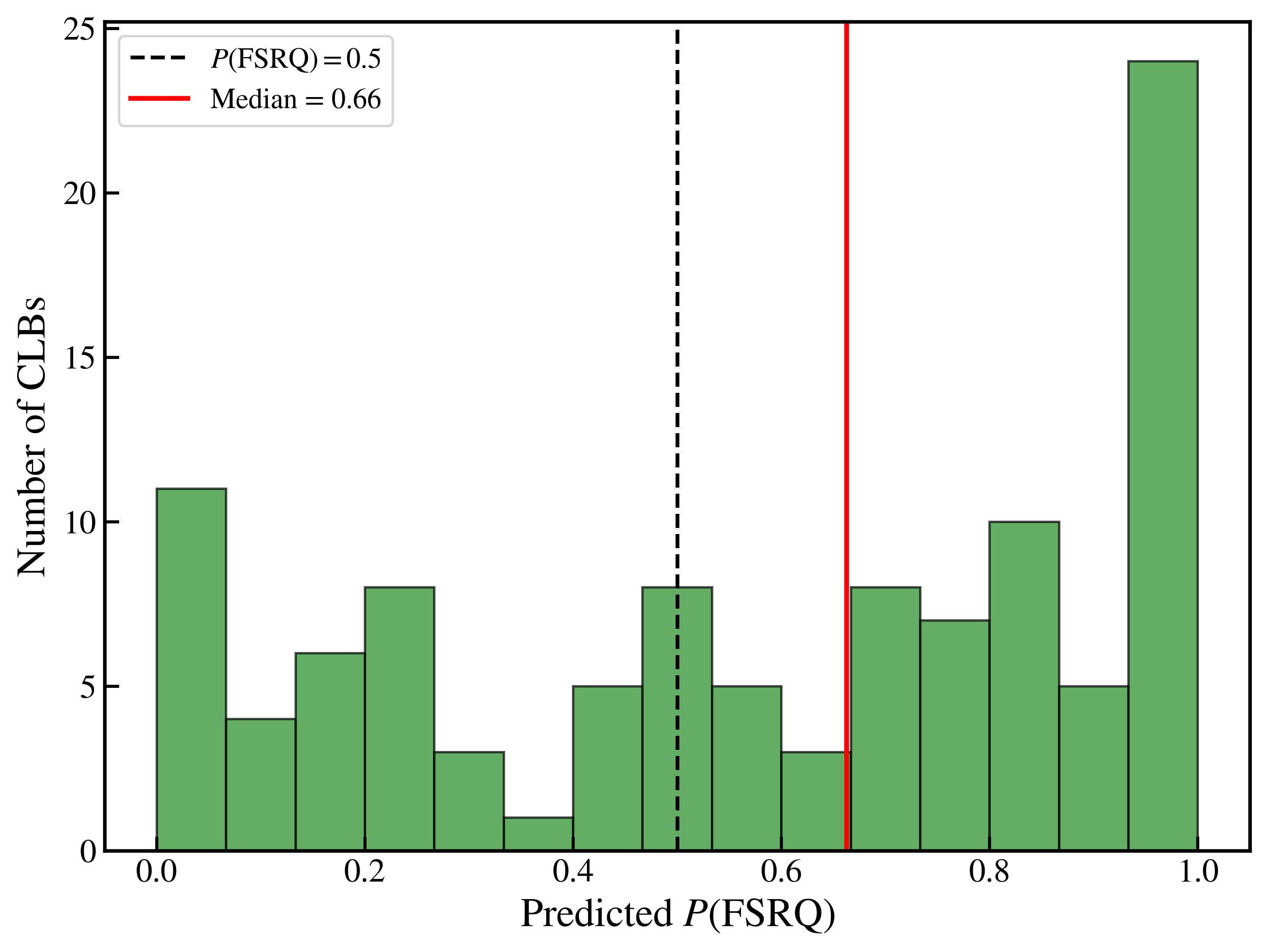}
\caption{
Distribution of the random-forest classification probabilities for the CLB population. 
}
\label{fig:rf_prob}
\end{figure*}

The resulting probability distribution is shown in Figure~\ref{fig:rf_prob}. The CLB population exhibits a broad range of predicted probabilities extending from the BLL-like to the FSRQ-like regime, indicating that transitional blazars do not form a completely homogeneous population. However, the overall distribution is shifted toward higher $P({\rm FSRQ})$ values, with a median probability of 0.66. \revone{A majority of CLBs are assigned a higher similarity to the FSRQ training population}, with 64 sources having $P({\rm FSRQ}) > 0.5$, while 50 and 26 sources exceed probability thresholds of 0.7 and 0.9, respectively.

Overall, the RF classification results are consistent with the statistical distributions and the PCA/UMAP analysis, further indicating that although CLBs occupy intermediate regions between the canonical subclasses, their overall population properties are comparatively more closely associated with the FSRQ population.

\section{X-ray Properties and X-ray/$\gamma$-ray Coupling}
\label{sec:xray}

\begin{figure*}
\centering

\includegraphics[width=0.85\textwidth]{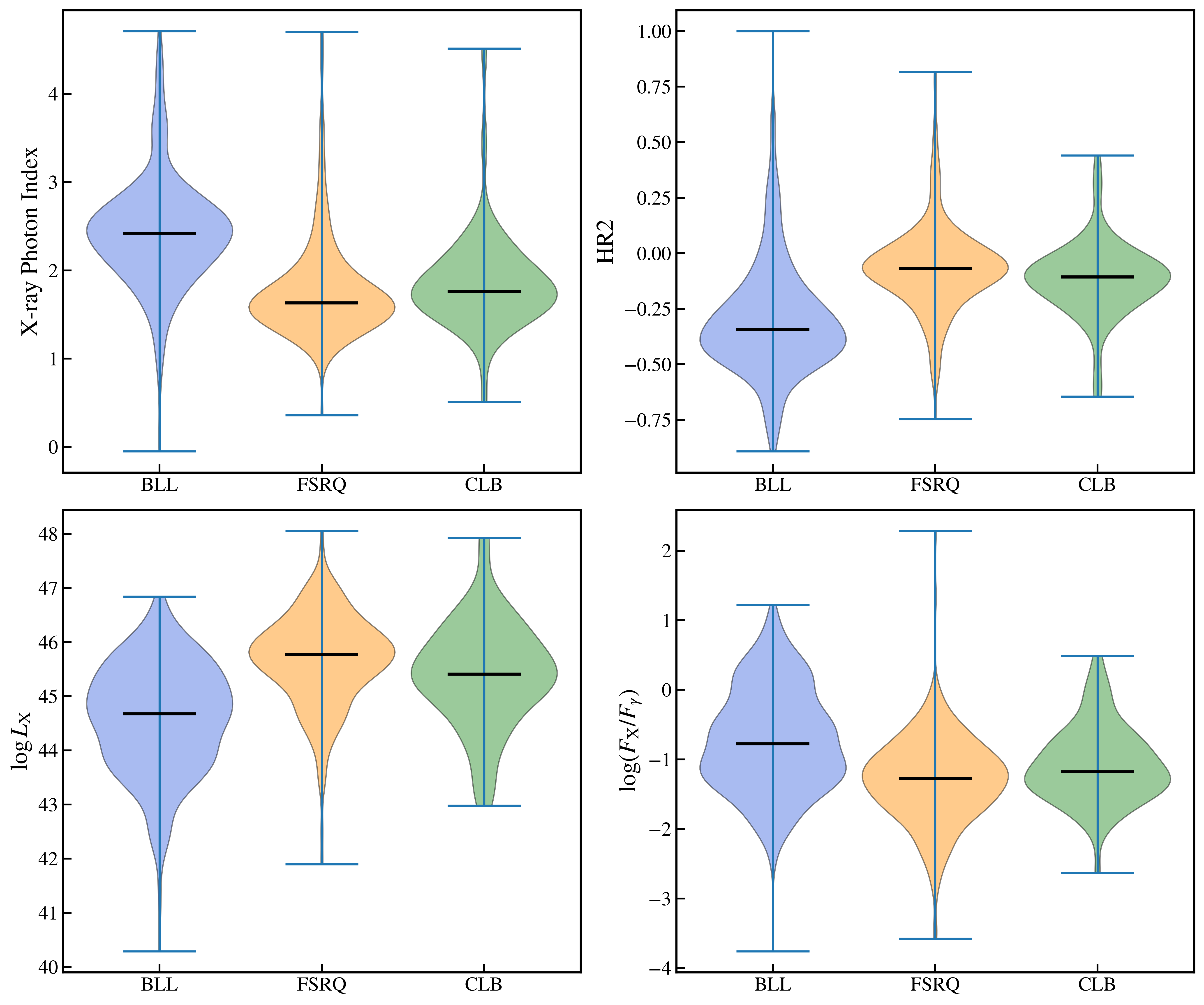}

\caption{
Violin distributions of the X-ray and X-ray/$\gamma$-ray properties for the LSXPS-associated CLB, BLL, and FSRQ samples. The figure shows the distributions of the X-ray photon index, HR2 hardness ratio, X-ray luminosity, and the $\log(F_{\rm X}/F_{\gamma})$ ratio for the three populations.}
\label{fig:xray_violin}
\end{figure*}

We further examined the X-ray properties of the LSXPS-associated sample described in Section~\ref{sec:data} using the X-ray photon index, HR2 hardness ratio, X-ray luminosity, and the X-ray/$\gamma$-ray flux and luminosity relations. The corresponding parameter distributions are shown in Figure~\ref{fig:xray_violin}.

The X-ray photon-index distributions show a noticeable separation between BLLs and FSRQs. BLLs generally occupy softer X-ray spectral regimes with larger photon-index values, whereas FSRQs are concentrated toward comparatively harder spectra. The CLB population mainly lies between the two groups, although the distribution is shifted more toward the FSRQ side. A similar behavior is observed for the HR2 hardness ratio, where BLLs exhibit systematically lower and more negative HR2 values while FSRQs remain clustered closer to zero. The CLB distribution again occupies the intermediate region but overall remains closer to the FSRQ population.

The lower panels of Figure~\ref{fig:xray_violin} present the distributions of the X-ray luminosity and the $\log(F_{\rm X}/F_{\gamma})$ ratio. The X-ray luminosities of CLBs span the region between BLLs and FSRQs, with the overall distribution extending closer toward the FSRQ regime. The $\log(F_{\rm X}/F_{\gamma})$ distributions also show that CLBs occupy intermediate regions between the two populations, indicating that the relative balance between X-ray and $\gamma$-ray emission in transitional blazars differs from that of typical BLLs, while the distinction from FSRQs is comparatively weaker.

The statistical differences between the populations were quantified using two-sample Kolmogorov--Smirnov (KS) tests, and the results are summarized in Table~\ref{tab:ks_xray}. The X-ray photon index and HR2 hardness ratio show substantially larger separation between CLBs and BLLs than between CLBs and FSRQs. Similar trends are observed for the X-ray luminosities and the X-ray/$\gamma$-ray flux and luminosity ratios, where the CLB--FSRQ comparisons consistently yield smaller KS statistics. \revone{As discussed in Section~\ref{sec:data} (Figure~\ref{fig:xray_matching_redshift}), the LSXPS-associated sample does not exhibit a significant redshift-dependent matching bias for either the CLB or FSRQ populations. Therefore, the closer CLB--FSRQ association observed in the X-ray parameter space is unlikely to be driven primarily by differential X-ray sample selection.} Overall, the X-ray properties indicate that the CLB population is statistically more closely associated with FSRQs than with BLLs.


\begin{table*}
\centering
\caption{Kolmogorov--Smirnov (KS) test results for the X-ray and broadband X-ray/$\gamma$-ray quantities derived from the LSXPS-associated sample.}
\label{tab:ks_xray}
\begin{tabular}{lcccc}
\hline
Parameter & KS$_{\rm CLB-BLL}$ & $p$-value & KS$_{\rm CLB-FSRQ}$ & $p$-value \\
\hline
X-ray Photon Index & 0.572 & $3.35\times10^{-13}$ & 0.236 & $3.35\times10^{-2}$ \\
HR2 Hardness Ratio & 0.556 & $1.23\times10^{-17}$ & 0.190 & $4.24\times10^{-2}$ \\
$\log L_{\rm X}$ & 0.412 & $1.46\times10^{-8}$ & 0.222 & $1.30\times10^{-2}$ \\
$\log(F_{\rm X}/F_{\gamma})$ & 0.270 & $2.61\times10^{-4}$ & 0.186 & $5.09\times10^{-2}$ \\
$\log(L_{\rm X}/L_{\gamma})$ & 0.346 & $3.89\times10^{-6}$ & 0.192 & $4.65\times10^{-2}$ \\
\hline
\end{tabular}
\end{table*}

\begin{figure*}
\centering

\includegraphics[width=0.85\textwidth]{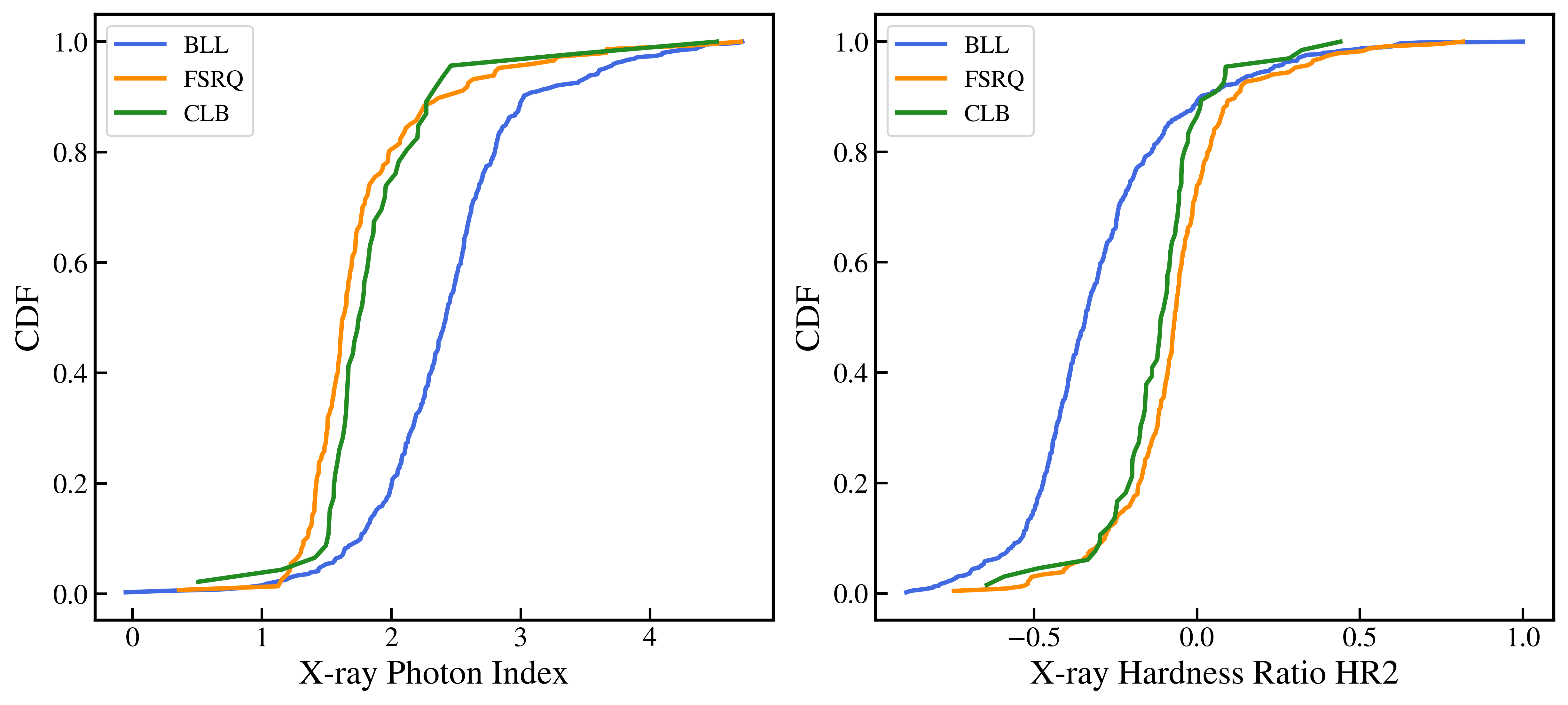}

\caption{
Cumulative distribution functions (CDFs) of the X-ray photon index and HR2 hardness ratio for the LSXPS-associated CLB, BLL, and FSRQ populations. 
}
\label{fig:xray_cdf}
\end{figure*}

The cumulative distribution functions shown in Figure~\ref{fig:xray_cdf} further support these trends. In both the X-ray photon index and HR2 distributions, the CLB population lies between the BLL and FSRQ populations, while remaining systematically closer to the FSRQ distributions over most of the cumulative range.

\begin{figure*}
\centering
\includegraphics[width=\textwidth]{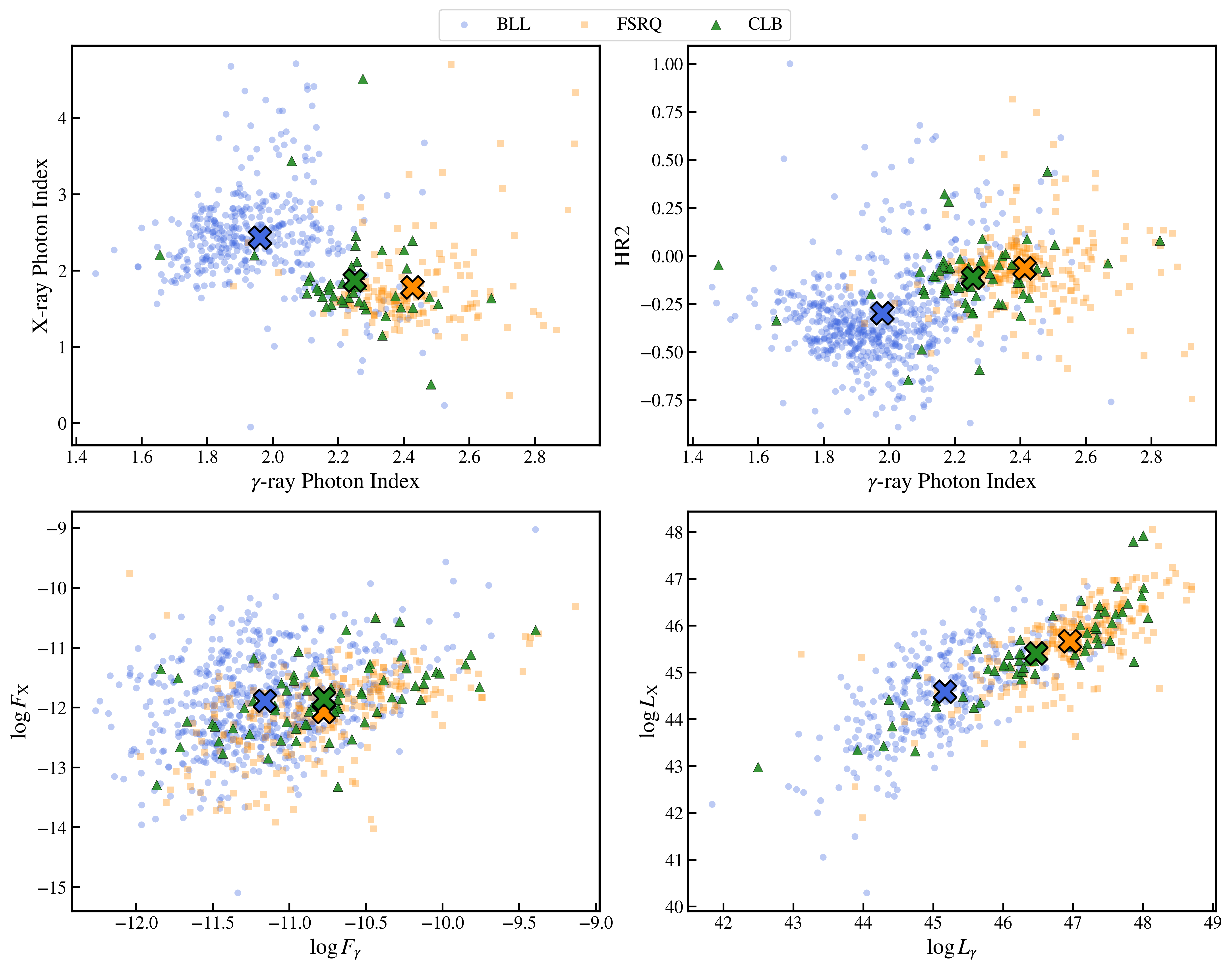}
\caption{
X-ray/$\gamma$-ray coupling relations for the BLL, FSRQ, and CLB populations. 
The panels show the relations between the $\gamma$-ray photon index and X-ray photon index (top-left), $\gamma$-ray photon index and HR2 hardness ratio (top-right), $\gamma$-ray and X-ray fluxes (bottom-left), and $\gamma$-ray and X-ray luminosities (bottom-right). 
The centroid markers indicate the \revtwo{mean} positions of the three subclasses in each parameter space. 
}
\label{fig:xray_gamma_coupling}
\end{figure*}

The X-ray/$\gamma$-ray coupling relations are presented in Figure~\ref{fig:xray_gamma_coupling}. The upper panels compare the $\gamma$-ray photon index with the X-ray photon index and HR2 hardness ratio. In both parameter spaces, the CLB population mainly occupies the overlap region between BLLs and FSRQs, but a substantial fraction of sources clusters closer to the FSRQ-dominated region. The centroid positions are also shifted toward the FSRQ side, consistent with the trends observed in the previous analyses. The lower panels show the relations between the X-ray and $\gamma$-ray fluxes and luminosities. A positive correlation is observed between the X-ray and $\gamma$-ray emission, particularly in the luminosity relation where the three populations form a comparatively tighter sequence. The CLB population again occupies intermediate regions between BLLs and FSRQs, while the overall distribution and centroid behavior remain comparatively closer to the FSRQ population.

The corresponding Spearman rank correlation coefficients are listed in Table~\ref{tab:xray_gamma_corr}. Among the explored relations, the $\log L_{\gamma}$--$\log L_{\rm X}$ relation shows the largest correlation coefficient for the CLB population with $\rho = 0.883$, while the $\log F_{\gamma}$--$\log F_{\rm X}$ relation also exhibits a positive correlation with $\rho = 0.526$. The flux and luminosity relations therefore indicate a comparatively close connection between the X-ray and $\gamma$-ray emission behavior of CLBs. In contrast, the relations involving the X-ray photon index and HR2 hardness ratio show weaker correlations within the individual populations, suggesting that the spectral-shape relations exhibit comparatively larger scatter than the flux and luminosity coupling relations.

Overall, the X-ray and X-ray/$\gamma$-ray coupling results support the picture that CLBs occupy transitional regions between BLLs and FSRQs while exhibiting X-ray and multiwavelength properties that remain comparatively closer to the FSRQ population.

\begin{table*}
\centering
\caption{Spearman rank correlation coefficients for the principal X-ray/$\gamma$-ray coupling relations. The table summarizes the broadband correlations for the complete sample as well as for the individual BLL, FSRQ, and CLB subclasses.}
\label{tab:xray_gamma_corr}
\begin{tabular}{llcc}
\hline
Relation & Class & $\rho$ & $p$-value\\
\hline

$\Gamma_{\gamma}$ vs X-ray Photon Index 
& All  & $-0.462$ & $3.37\times10^{-32}$\\
& BLL  & $-0.020$ & $6.94\times10^{-1}$ \\
& FSRQ & $0.044$  & $5.96\times10^{-1}$ \\
& CLB  & $-0.256$ & $8.59\times10^{-2}$ \\
\hline

$\Gamma_{\gamma}$ vs HR2
& All  & $0.517$ & $1.63\times10^{-61}$  \\
& BLL  & $0.205$ & $5.87\times10^{-7}$ \\
& FSRQ & $0.104$ & $1.12\times10^{-1}$  \\
& CLB  & $0.246$ & $4.68\times10^{-2}$  \\
\hline

$\log F_{\gamma}$ vs $\log F_{\rm X}$
& All  & $0.311$ & $3.47\times10^{-21}$  \\
& BLL  & $0.283$ & $3.41\times10^{-12}$  \\
& FSRQ & $0.545$ & $2.71\times10^{-19}$  \\
& CLB  & $0.526$ & $5.65\times10^{-6}$ \\
\hline

$\log L_{\gamma}$ vs $\log L_{\rm X}$
& All  & $0.799$ & $1.57\times10^{-133}$  \\
& BLL  & $0.697$ & $2.47\times10^{-47}$  \\
& FSRQ & $0.733$ & $5.37\times10^{-38}$  \\
& CLB  & $0.883$ & $1.01\times10^{-21}$  \\
\hline

\end{tabular}
\end{table*}

\section{Summary and Discussion}
\label{sec:discussion}

The main results of this work can be summarized as follows:

\begin{itemize}

\item CLBs mainly occupy intermediate and overlapping regions between the BLL and FSRQ populations in both the $\gamma$-ray and X-ray parameter spaces.

\item The PCA and UMAP projections show that CLBs populate the transition region connecting the two main blazar populations rather than forming a completely isolated group. \revtwo{An unsupervised Gaussian mixture analysis of the same parameter space likewise recovers no distinct CLB-dominated component (\ref{appendix:gmm}).}

\item The statistical distributions, centroid trends, and random-forest classification probabilities consistently indicate that the overall properties of CLBs remain comparatively closer to the FSRQ population. The variability index distributions are statistically indistinguishable between CLBs and FSRQs (KS$_{\rm CLB-FSRQ}$ = 0.074, $p$ = 0.648), in contrast to the larger separation observed between CLBs and BLLs (KS$_{\rm CLB-BLL}$ = 0.445, $p$ = $1.47\times10^{-18}$). This result indicates that, on average, the long-term $\gamma$-ray variability properties of CLBs are more similar to those of FSRQs than to BLLs.

\item \revone{The X-ray and X-ray/$\gamma$-ray coupling analysis shows similar behavior, with the CLB population occupying intermediate regions while remaining statistically closer to FSRQs across all examined X-ray parameters. The cumulative distribution analysis (Figure~\ref{fig:xray_cdf}) further reinforces this result, showing that the CLB population remains systematically closer to the FSRQ
population than to BLLs across nearly the entire cumulative range of the X-ray photon index and HR2 hardness ratio.}

\item The X-ray and $\gamma$-ray luminosities exhibit positive correlations for all populations. The $\log L_{\gamma}$--$\log L_{\rm X}$ relation for CLBs yields $\rho=0.883$, exceeding the corresponding values for BLLs ($\rho=0.697$) and FSRQs ($\rho=0.733$), suggesting that the X-ray and $\gamma$-ray emission processes in transitional blazars are driven more coherently than in either canonical subclass.

\end{itemize}

The overall behavior of CLBs observed in this work is broadly consistent with earlier studies suggesting that transitional blazars may represent systems located near the boundary between BLLs and FSRQs \citep{2012MNRAS.420.2899G,2024ApJ...962..122K}. In the standard blazar picture, the observed differences between the two established subclasses are commonly linked to differences in accretion efficiency, radiative cooling, and the strength of the external radiation environment \citep{1998MNRAS.299..433F,1998MNRAS.301..451G}. FSRQs are generally associated with radiatively efficient accretion disks and strong external photon fields, whereas BLLs are thought to be associated with radiatively inefficient accretion flows and comparatively weaker external radiation environments \citep{1998MNRAS.299..433F,2011ApJ...740...98M,2015ApJ...810...14A,2020ApJ...892..105A}.

The statistical comparisons obtained in this work consistently suggest that the presently known CLB population is not distributed symmetrically between BLLs and FSRQs, but instead remains preferentially shifted toward the FSRQ population. In the $\gamma$-ray parameter space, the CLB--FSRQ KS statistics for the variability index and pivot energy are significantly smaller than the corresponding CLB--BLL values, with KS$_{\rm CLB-FSRQ}=0.074$ and $0.269$ compared to KS$_{\rm CLB-BLL}=0.445$ and $0.527$, respectively. Notably, the variability index distribution of CLBs is statistically indistinguishable from that of FSRQs ($p$ = 0.648), indicating that the long-term $\gamma$-ray variability properties of the CLB population are more similar to those of FSRQs than to BLLs. Since the variability index is derived from the time-averaged 4FGL-DR4 catalog, this result should be interpreted as a population-level property rather than evidence for the variability behavior during individual changing-look states. A similar trend is observed in the X-ray analysis, where the CLB--FSRQ KS statistics for the X-ray photon index, HR2 hardness ratio, and X-ray luminosity are 0.236, 0.190, and 0.222, whereas the corresponding CLB--BLL values increase to 0.572, 0.556, and 0.412. The X-ray/$\gamma$-ray flux and luminosity ratios also follow the same behavior, with KS$_{\rm CLB-FSRQ}=0.186$ and $0.192$ compared to KS$_{\rm CLB-BLL}=0.270$ and $0.346$. The multidimensional PCA and UMAP projections together with the random-forest classification analysis further support this behavior, where the median predicted FSRQ probability for CLBs reaches 0.66, while 64 out of 108 CLBs are assigned $P({\rm FSRQ})>0.5$ and 50 sources exceed $P({\rm FSRQ})>0.7$.

\revone{Among the CLB population, 26 sources are assigned $P(\mathrm{FSRQ}) > 0.9$, indicating that these objects occupy regions of the multidimensional $\gamma$-ray parameter space that are strongly associated with the FSRQ training population. Since the random-forest classifier was trained exclusively on the two canonical subclasses, these probabilities should not be interpreted as calibrated class-membership probabilities, but rather as indicators of similarity to the parameter space occupied by the BLL and FSRQ training samples. The predominance of high $P(\mathrm{FSRQ})$ values nevertheless suggests that a substantial fraction of CLBs share $\gamma$-ray properties more closely resembling those of FSRQs than BLLs, consistent with the independent trends revealed by the statistical distributions and the PCA/UMAP analyses. Within this interpretation, these sources may correspond to CLBs in which enhanced jet emission dominates the observed high-energy characteristics, consistent with scenarios in which strong non-thermal continuum emission temporarily suppresses or dilutes the broad-emission-line signatures that would otherwise identify these objects as FSRQs \citep{2012MNRAS.420.2899G}. Rather than providing definitive subclass assignments, the random-forest results therefore support the broader picture that CLBs occupy a transitional region in parameter space while exhibiting overall high-energy properties that are preferentially associated with the FSRQ population.}

The correlation analysis provides additional insight into the high-energy emission properties of transitional blazars. The comparatively tight $\log L_{\gamma}$--$\log L_{\rm X}$ relation observed for the CLB population ($\rho=0.883$) suggests that the X-ray and $\gamma$-ray emission processes in CLBs remain closely and coherently connected. The corresponding correlation is weaker for the BLL and FSRQ populations, with $\rho=0.697$ and $0.733$, respectively. In leptonic blazar models, such tight multiwavelength coupling naturally arises when a single electron population dominates both the X-ray and $\gamma$-ray emission through related radiative processes. The tighter coupling observed in CLBs relative to BLLs therefore suggests that the jet radiative environment in transitional blazars remains more FSRQ-like at the high-energy emission level, even during periods when the optical classification shifts toward the BLL regime. Similarly, the $\log F_{\gamma}$--$\log F_{\rm X}$ relation for CLBs shows a positive correlation with $\rho=0.526$, comparable to the FSRQ population and noticeably stronger than the BLL value of $\rho=0.283$.

The luminosity comparisons in this work are considered alongside the known redshift distributions of the three blazar subclasses. FSRQs in \textit{Fermi} catalogs are systematically found at higher redshifts than BLLs, and CLBs occupy an intermediate redshift range, with mean redshifts of approximately $z\sim1.2$, $z\sim0.9$, and $z\sim0.4$ for FSRQs, CLBs, and BLLs respectively \citep{2024ApJ...962..122K}. In flux-limited surveys such as 4FGL-DR4, luminosity distributions can partly be influenced by the common dependence on redshift. However, the FSRQ bias observed in this work is not confined to luminosity comparisons alone. The spectral indices, hardness ratios, and variability properties, which are independent of distance, all consistently show that CLBs are statistically closer to FSRQs than to BLLs, with KS statistics smaller by factors of two to six across the explored parameter space. The multidimensional PCA and UMAP projections, which use only distance-independent spectral and variability parameters, reproduce the same FSRQ bias in the CLB centroid positions. The overall conclusion that CLBs are physically closer to FSRQs therefore remains independent of differential redshift distributions and is unlikely to be purely a distance-induced effect. \revone{This conclusion is further supported by the redshift-dependent matching analysis presented in Figure~\ref{fig:xray_matching_redshift}, which shows that the X-ray matching probability itself does not exhibit a significant dependence on redshift for either the CLB or FSRQ populations.} The intermediate redshift distribution of CLBs is itself consistent with the interpretation that transitional blazars may represent an evolutionary phase between FSRQs and BLLs across cosmic time \citep{2014ApJ...797...19R,2024ApJ...962..122K}.

\revone{Although the present analysis is based on population statistics rather than evolutionary modeling, the observed redshift distribution together with the high-energy properties of the three blazar populations is broadly consistent with a possible evolutionary sequence linking FSRQs and BLLs. \revtwo{An evolutionary connection of this kind has been proposed previously. \citet{2002ApJ...571..226C} and \citet{2002ApJ...564...86B} argued that the blazar spectral sequence also traces an evolutionary one, in which gas-rich, radiatively efficient FSRQs evolve into gas-depleted BL Lac objects as the accretion rate declines, and \citet{2003ApJ...593..667M} reached a similar conclusion from the jet--disk connection. Observational support was subsequently provided by \citet{2014ApJ...780...73A}, who found the cosmic evolution of the \textit{Fermi} BL Lac population to be consistent with such a link, with the space density of BL Lacs rising at the epoch where the FSRQ population declines. Within this framework, changing-look blazars would occupy the transitional stage.} In this speculative picture, powerful FSRQs with radiatively efficient accretion may evolve through a transitional changing-look phase as the accretion rate decreases and the relative contribution of the non-thermal jet emission increases, eventually giving rise to low-synchrotron-peaked BL Lac objects and subsequently to intermediate- and high-synchrotron-peaked BL Lacs. As the accretion rate decreases, the associated external photon fields become weaker, reducing external-Compton cooling and potentially allowing electrons to be accelerated to higher energies, thereby shifting the synchrotron peak toward the ISP and HSP regimes. Such an interpretation is qualitatively consistent with the intermediate redshift distribution of CLBs, their intermediate $\gamma$-ray properties, and their closer overall association with the FSRQ population found in both the $\gamma$-ray and X-ray analyses. Nevertheless, the present work does not establish such an evolutionary sequence, and dedicated long-term multiwavelength studies tracing the temporal evolution of individual changing-look blazars will be required to test this scenario.}

The multidimensional distributions suggest that the separation between the two main blazar subclasses may not be completely discrete, with the CLB population occupying a continuous transition region between the BLL and FSRQ distributions. Such behavior suggests that at least a fraction of blazars may pass through intermediate observational states that are not fully represented within the traditional optical classification scheme. The high-energy behavior of CLBs may therefore be linked to physical conditions commonly associated with FSRQs. In leptonic blazar models, the strong external photon fields produced by the accretion disk, broad-line region, and dusty torus in FSRQs can enhance external-Compton emission and produce luminous and comparatively softer high-energy spectra \citep{1998MNRAS.301..451G,2011ApJ...740...98M}. In contrast, BLLs are generally associated with weaker external radiation environments and synchrotron self-Compton dominated emission. If many CLBs still retain partially efficient accretion environments, their X-ray and $\gamma$-ray properties would naturally remain more similar to FSRQs even during phases in which their optical spectra appear BLL-like. This interpretation is also consistent with scenarios in which changing-look behavior is driven by variations in accretion activity or temporary dilution of broad emission lines by enhanced jet continuum emission \citep{2012MNRAS.420.2899G,2012MNRAS.425.1371G}.

\revone{One limitation of the present study is that the $\gamma$-ray parameters analyzed here (e.g., photon index, pivot energy, variability index, and spectral curvature) are adopted from the time-averaged 4FGL-DR4 catalog and therefore represent the long-term average high-energy properties of each source over the \textit{Fermi}-LAT observing period. Consequently, these parameters do not capture the short-term spectral and flux evolution that characterizes the changing-look phenomenon itself. The present work is therefore intended as a population-level investigation of the average high-energy properties of CLBs rather than a time-resolved study of individual state transitions. Future epoch-resolved analyses of individual CLBs, based on separate \textit{Fermi}-LAT time intervals and coordinated with optical changing-look episodes where available, will provide a more direct probe of the temporal evolution associated with the changing-look phenomenon.}

\revtwo{A closely related limitation concerns the optical state occupied by each
CLB during the \textit{Fermi}-LAT observing period itself. Sources enter the
changing-look catalog on the basis of optical spectra obtained at epochs that are
neither contemporaneous with nor uniformly distributed across the LAT mission,
and the catalog does not record the fraction of the LAT exposure that each source
spent in an FSRQ-like or a BLL-like state. A source that occupied an FSRQ-like
state for most of the mission would be expected to exhibit FSRQ-like
time-averaged high-energy properties, so the population-level proximity to FSRQs
reported here may partly reflect the state occupancy of the present sample rather
than a property intrinsic to transitional blazars. We note that the sample is not
composed predominantly of sources carrying an FSRQ optical classification: of the 109
CLBs considered here, 61 (56 per cent) are classified as \texttt{bll} and 44 (40
per cent) as \texttt{fsrq} in 4FGL-DR4. A quantitative treatment of this effect
nevertheless requires epoch-resolved $\gamma$-ray analysis in which the LAT data
are segmented according to documented optical states and compared with
contemporaneous optical spectroscopy, which lies beyond the scope of the present
population-level study. We identify this as an important direction for future
work on the changing-look blazar population.}

\revone{Another potential limitation concerns the interpretation of the \textit{Fermi}-LAT Variability Index. Although the Variability Index is independent of redshift by definition, it is influenced by photon statistics, such that sources with fewer detected photons generally have larger uncertainties in the estimated variability. Since FSRQs are, on average, located at higher redshifts than BLLs, such sensitivity effects could moderately influence the statistical separation in the Variability Index parameter space. However, the principal conclusions of this work do not rely on the Variability Index alone, but are consistently supported by the combined behavior of the photon index, pivot energy, spectral curvature, multidimensional analyses (PCA and UMAP), random-forest classification, and the independent X-ray population analysis.}

The presently known CLB population remains comparatively small relative to the large BLL and FSRQ samples, and the identification of changing-look behavior still depends strongly on long-term spectroscopic and multiwavelength observations. Future optical, X-ray, and $\gamma$-ray monitoring together with larger samples of confirmed transitional systems will therefore be important for understanding the origin and long-term evolution of changing-look blazars.

\section*{Acknowledgements}

We thank the anonymous referee for the careful reading of the manuscript and for the constructive comments and suggestions, which have significantly improved the quality and clarity of this work. The authors acknowledge the use of public data and software provided by the \textit{Fermi} Science Support Center (FSSC). This work has made use of the \textit{Fermi}-LAT Fourth Source Catalog Data Release 4 (4FGL-DR4), generated using observations from the \textit{Fermi} Large Area Telescope mission. This research has also made use of the LSXPS catalog derived from observations obtained with the Neil Gehrels \textit{Swift} Observatory. This research has made use of the CLBCat online source catalog (\url{https://github.com/ksj7924/CLBCat}). ZM acknowledges the financial support provided by the Science and Engineering Research Board (SERB), Government of India, under the National Postdoctoral Fellowship (NPDF), Fellowship reference no. PDF/2023/002995. ZS is supported by the Department of Science and Technology, Govt. of India, under the INSPIRE Faculty grant (DST/INSPIRE/04/2020/002319).


\bibliographystyle{elsarticle-harv} 
\bibliography{sample631}





 
\appendix

\section{HR1 Hardness-Ratio Analysis}
\label{appendix:hr1}

\revone{To assess whether the selection of HR2 as the principal X-ray hardness-ratio diagnostic could introduce selective-reporting bias, we repeated the population analysis using the LSXPS HR1 hardness ratio.}

\begin{figure}[ht]
\centering
\includegraphics[width=0.75\linewidth]
{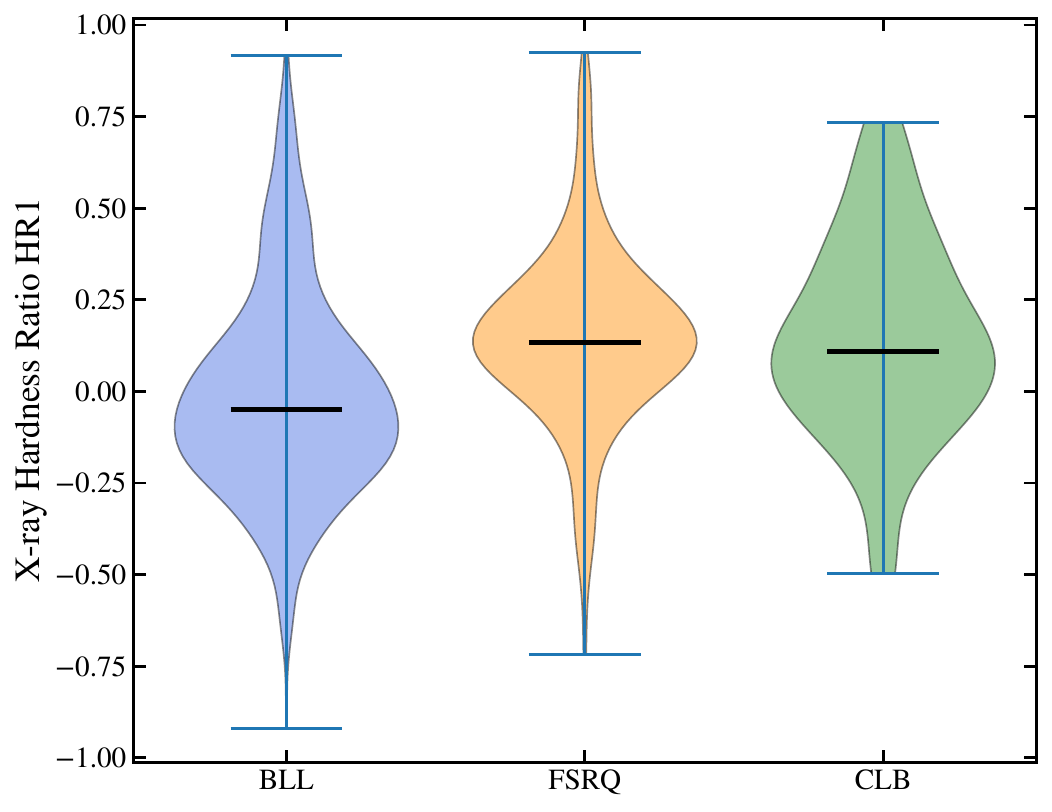}
\caption{Distribution of the LSXPS HR1 hardness ratio for the BLL, FSRQ, and CLB populations. The horizontal black bars denote the sample medians.}
\label{fig:hr1_violin}
\end{figure}

\begin{figure}[ht]
\centering
\includegraphics[width=0.75\linewidth]
{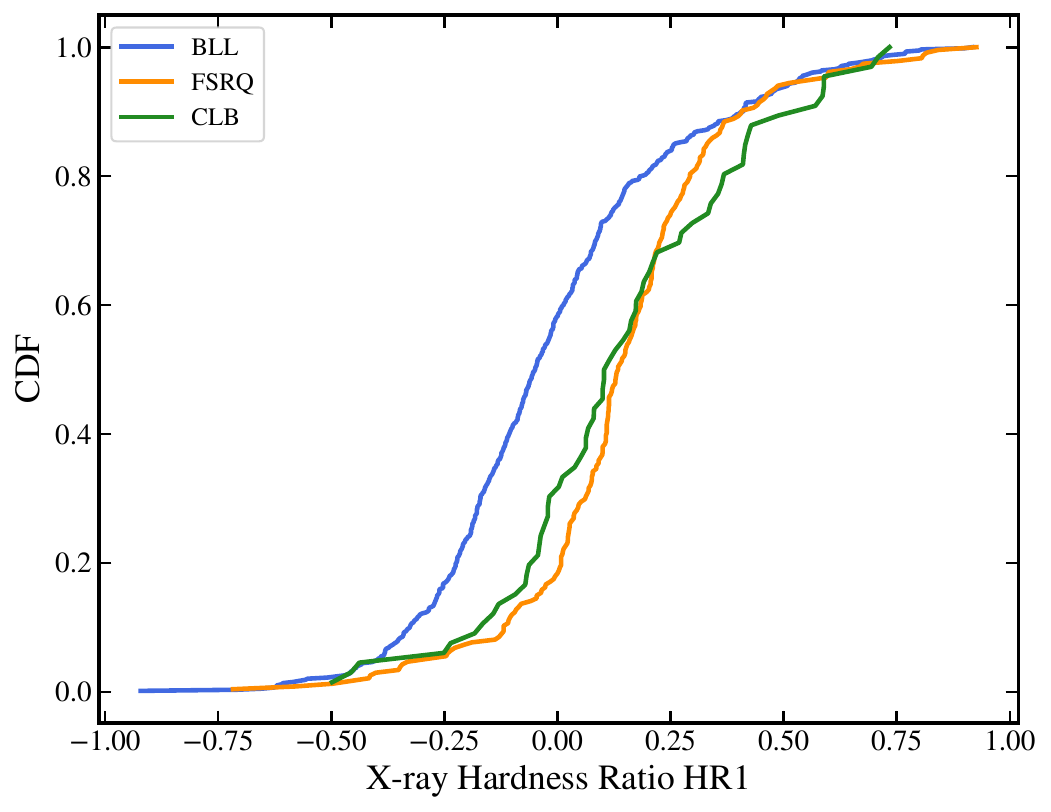}
\caption{Cumulative distribution functions of the LSXPS HR1 hardness ratio for the BLL, FSRQ, and CLB populations.}
\label{fig:hr1_cdf}
\end{figure}

\revone{The resulting HR1 distributions and cumulative distribution functions are shown in Figures~\ref{fig:hr1_violin} and \ref{fig:hr1_cdf}, respectively. The corresponding two-sample Kolmogorov--Smirnov test results are summarized in Table~\ref{tab:hr1_ks}.}

\revone{The HR1 analysis yields the same qualitative conclusion as the HR2 analysis. The CLB population differs significantly from the BLL population (KS = 0.318, $p=8.3\times10^{-6}$), whereas no statistically significant difference is found between the CLB and FSRQ populations (KS = 0.136, $p=0.266$). These results independently support the conclusion that the X-ray properties of CLBs are overall more closely associated with the FSRQ population. Nevertheless, HR2 was retained in the main analysis because it provides comparatively clearer subclass separation and more stable statistical behavior.}

\begin{table}[ht]
\centering
\caption{Kolmogorov--Smirnov (KS) test results for the HR1 hardness-ratio distributions. Similar to the HR2 analysis presented in the main text, the HR1 results indicate that the CLB population differs significantly from the BLL population, while no statistically significant difference is found between the CLB and FSRQ populations.}
\label{tab:hr1_ks}
\begin{tabular}{lcc}
\hline
Comparison & KS statistic & $p$-value \\
\hline
CLB--BLL  & 0.318 & $8.27 \times 10^{-6}$ \\
CLB--FSRQ & 0.136 & 0.266 \\
\hline
\end{tabular}
\end{table}

\section{Unsupervised Gaussian Mixture Analysis}
\label{appendix:gmm}

\revtwo{To test whether CLBs form a distinct group in the multidimensional
$\gamma$-ray parameter space, or are instead distributed among the regions
occupied by the canonical subclasses, we performed an unsupervised Gaussian
mixture model (GMM) analysis. The analysis used the same four standardized
$\gamma$-ray parameters and the same analysis sample (108 CLBs, 1429 BLLs, and
775 FSRQs) employed in the PCA, UMAP, and random-forest analyses. A
three-component mixture was fitted without reference to the source
classifications, with the covariance parameterization selected by the Bayesian
information criterion, which favored the full-covariance model.}

\begin{figure*}[ht]
\centering
\includegraphics[width=0.9\textwidth]{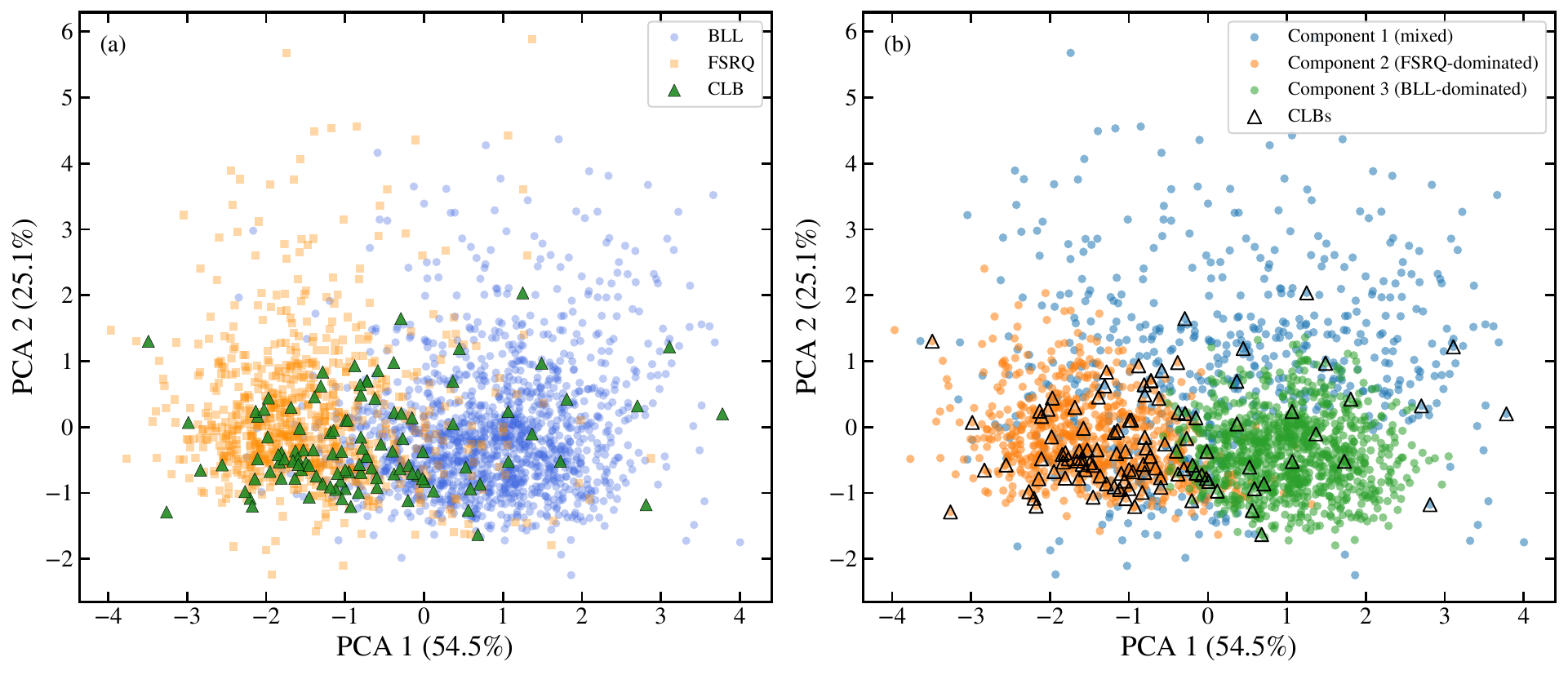}
\caption{\revtwo{Unsupervised three-component Gaussian mixture model of the
$\gamma$-ray analysis sample, shown in the two-dimensional PCA projection. Panel
(a) shows the sources colored by their catalog classification, and panel (b) by
the mixture component to which they are assigned. No CLB-dominated component is
recovered, and the CLBs are distributed among the components occupied by the
canonical BLL and FSRQ populations.}}
\label{fig:gmm}
\end{figure*}

\revtwo{The resulting component structure is shown in
Figure~\ref{fig:gmm}. No CLB-dominated component is recovered: the largest CLB
fraction within any component is 9.0 per cent. Instead, 62.0 per cent (67/108) of
the CLBs are associated with the predominantly FSRQ component, 22.2 per cent
(24/108) with the predominantly BLL component, and the remaining 15.7 per cent
(17/108) with a mixed component. Repeating the complete fit for 50 independent
random initializations left the selected covariance parameterization unchanged,
produced successful convergence in every realization, and recovered no
CLB-dominated component in any of them. The result is therefore insensitive to
the choice of random initialization.}

\revtwo{This behavior is consistent with previous population studies.
\citet{2024ApJ...962..122K} found the physical properties of CLBs to lie between
those of BLLs and FSRQs, and \citet{10.1093/mnras/stag542} reached the same
conclusion from the $\gamma$-ray photon spectral index. It is also consistent
with the more general picture in which the boundary between the two canonical
subclasses is continuous rather than discrete
\citep{2011ApJ...740...98M,2012MNRAS.420.2899G}. We note that
\citet{2025ApJ...980..213K} recovered a three-group mixture solution that
includes a CLB-candidate group. The parameter combinations favored in that
analysis are, however, dominated by accretion-related quantities such as the
Eddington-scaled disk luminosity, the disk luminosity, and the Compton dominance
rather than by $\gamma$-ray catalog observables, and the reported agreement
between the recovered partition and the optical classification is modest. Their
candidate group is likewise located between the two canonical populations and is
dominated by sources carrying an FSRQ classification in the 4FGL catalog. The
present result is therefore consistent in direction with that work, while
indicating that within the four-dimensional $\gamma$-ray observable space alone
the CLB population is not separable as an independent component.}

\revtwo{Such behavior is physically expected. The optical classification rests
on a threshold in emission-line equivalent width applied to what is a continuous
distribution of broad-line strength relative to the beamed jet continuum. Whether
a source falls on one side of that threshold depends on the relative
contributions of the accretion-powered external radiation field and the
non-thermal jet continuum, both of which vary continuously across the population.
A discrete boundary in the optical classification therefore need not correspond
to a distinct component in the high-energy observable space, and transitional
sources are expected to populate the intermediate region rather than to cluster
separately.}

\end{document}